\documentclass[aps,pra,twocolumn,amsmath,amssymb,showpacs,superscriptaddress]{revtex4-1}
\usepackage{xcolor}
\usepackage{textcomp} 
\usepackage{mathrsfs,amsmath}
\usepackage{graphicx}
\usepackage{dcolumn}
\usepackage[mathlines]{lineno}
\usepackage{hyperref}
\usepackage{bm}
\usepackage{epstopdf}
\usepackage{soul}
\usepackage{epsfig}
\usepackage[normalem]{ulem}
\usepackage{bbold}
\usepackage{braket}
\usepackage{physics}
\usepackage{gensymb}
\usepackage{amsmath,amssymb}
\usepackage{soul}

\begin{document}
\raggedbottom
\title{Remote engineering of particle-like topologies to visualise entanglement dynamics} 

\author{Fazilah Nothlawala}
\affiliation{School of Physics, University of the Witwatersrand, Private Bag 3, Wits 2050, South Africa}

\author{Bereneice Sephton}
\email{bereneice.sephton@unina.it}
\affiliation{Dipartimento di Fisica, Universit\`a di Napoli Federico II, Complesso Universitario di Monte S. Angelo, Via Cintia, 80126 Napoli, Italy}

\author{Pedro Ornelas}
\affiliation{School of Physics, University of the Witwatersrand, Private Bag 3, Wits 2050, South Africa}

\author{Mwezi Koni}
\affiliation{School of Physics, University of the Witwatersrand, Private Bag 3, Wits 2050, South Africa}

\author{Bruno Piccirillo}
\affiliation{Dipartimento di Fisica, Universit\`a di Napoli Federico II, Complesso Universitario di Monte S. Angelo, Via Cintia, 80126 Napoli, Italy}

\author{Liang Feng}
\affiliation{Department of Electrical and Systems Engineering, University of Pennsylvania, Philadelphia, PA 19104, USA}

\author{Isaac Nape}
\affiliation{School of Physics, University of the Witwatersrand, Private Bag 3, Wits 2050, South Africa}

\author{Vincenzo D'Ambrosio}
\affiliation{Dipartimento di Fisica, Universit\`a di Napoli Federico II, Complesso Universitario di Monte S. Angelo, Via Cintia, 80126 Napoli, Italy}

\author{Andrew Forbes}
\affiliation{School of Physics, University of the Witwatersrand, Private Bag 3, Wits 2050, South Africa}

\begin{abstract}
\noindent \textbf{Skyrmions are a particle-like topology with a quantised skyrmion number, realised across condensed matter and photonic platforms alike. In quantum photonics, they constitute an emerging resource, promising robust quantum information encoding, so far realised as single photon and bi-photon entangled states. Here we report the first visualisation of tripartite entanglement dynamics through topological structure using spin-skyrmion entangled states, where the topology of a single photon is remotely controlled through the spin of its entangled partner. We visualise our tripartite state theoretically by introducing the notion of a topological Bloch sphere that completely captures the entanglement and topolological features of the state. By leveraging this state, we realise the first quantum multiskyrmions, comprising multiple localised skyrmions within a single structure, that emulate signatures of their magnetic counterparts. We verify this experimentally and show that traversing our topological sphere reveals entanglement-driven particle-like motion of the localised topological structures. These dynamics unveil a physical manifestation of tripartite entanglement correlations which we illustrate by example of GHZ-like states, enabling a visualisation of multiple Bell states encoded within our system. Our work opens exciting possibilities for quantum sensing by mapping complex quantum channel features onto topological observables of multipartite states and offers a promising avenue for harnessing quantum topologies for multi-level encoding quantum communication schemes.}
\end{abstract}



\maketitle

\noindent Originally introduced as solitonic solutions in nonlinear field theories \cite{skyrme1961non}, skyrmions are a particle-like topology characterized by an integer-valued topological charge, known as the skyrmion number ($n$), which is associated with the wrapping of a spin or pseudospin field. These topological excitations have since been realised in many systems, from magnetic materials \cite{nagaosa2013topological, fert2017magnetic}, ultracold atomic gases \cite{leslie2009creation,zhang2018topological}, twistronics \cite{kwan2022skyrmions}, liquid-crystals \cite{ackerman2014two,smalyukh2010three} and acoustics \cite{ge2021observation,hu2023observation} to optical fields \cite{shen2024optical,yang2025optical,lei2025topological}. Underpinned by topological stability, they form attractive candidates for the storage and transfer of information \cite{zhang2020skyrmion, han2022high, lima2022spin}.

\noindent Their optical implementation has been explored with surface plasmon polaritons \cite{bai2020dynamic,tsesses2018optical,davis2020ultrafast}, evanescent fields \cite{lin2021photonic}, spatiotemporal pulses \cite{teng2025construction,shen2021supertoroidal}, Poynting vectors \cite{wang2024topological} and Stokes parameters of structured paraxial beams \cite{gao2020paraxial,shen2022generation,cisowski2023building}. In the Stokes approach, the spatial phase and polarization are engineered to produce these spin-textured fields in free space, with the Stokes skyrmion defined as a mapping between the tranverse spatial plane, $\mathcal{R}^2$ and a polarization state space, $\mathcal{S}^2$, represented by the Poincare sphere. These optical Stokes skyrmions offer the advantage of a versatile platform for investigating exotic topologies \cite{cisowski2023building}, light-matter exchange \cite{mitra2025topological,stanciu2007all,lambert2014all} for future data storage devices \cite{fert2013skyrmions,foster2019two} and the promise of computing \cite{wang2025perturbation} as well as optical processing at the speed of light. 

\noindent Only recently have these structures been realised in the quantum regime, as non-local states across entangled photons \cite{ornelas2024non}, locally in single photons \cite{ma2025nanophotonic} and controllably in both regimes (local and non-local) \cite{koni2025dual}. The topology in these systems, however, has remained a well-defined feature of the underlying states from which they are derived. 

\noindent Here, we introduce spin-skyrmion entangled states, where the skyrmion topology of one photon is remotely controlled by its entangled twin, with the underlying topological structure revealing tripartite entanglement dynamics. We demonstrate these features through experimentally measured dual-wavelength entangled states which exhibit topological transitions, switching between two different skyrmion numbers. To visualise this skyrmion number transition, we introduce the notion of a topological Bloch sphere, capturing the state of the skyrmion-photon given measurements on the spin-photon. Notably, no well-defined spin texture or skyrmion number exists prior to measurement of the entangled spin-photon, underscoring the non-local character of the topology. This allows us to report the first creation of quantum multiskyrmions -- complex skyrmion structures characterized by multiple skyrmion numbers -- exhibiting remotely driven quasiparticle dynamics, emulating magnetic systems. Lastly, we isolate embedded GHZ-like states whose behaviour can be identified with the evolution of their topological spin-textures, thereby revealing a physical manifestation of complex tripartite entanglement dynamics. These results open new avenues for harnessing dynamic topological quasiparticles within quantum sensing protocols by encoding complex channel features onto topological observables, enabling physical observation of multipartite state evolution.

\begin{figure*}[!htbp]
    \centering
    \includegraphics[width=1\linewidth]{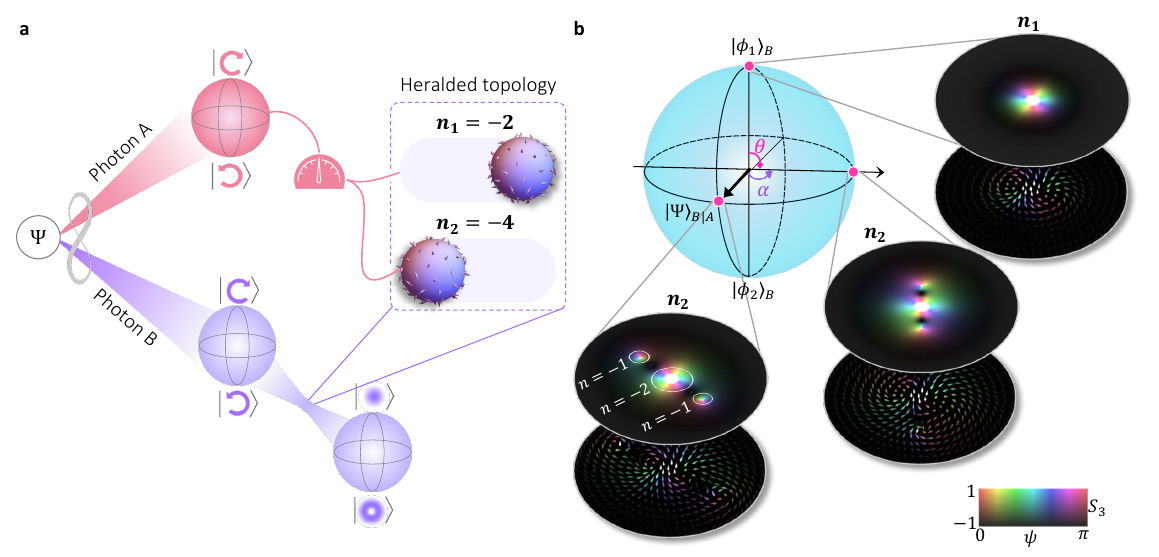}
    \caption{\textbf{Non-local topological control and creation of quantum multiskyrmions.} \textbf{a,} Photons A and B form an entangled state, where the polarization of photon A is coupled to a skyrmion state state on photon B. A measurement made on the polarization of photon A will collapse the state of photon B into a certain skyrmion state with a topological skyrmion number that can be switched between $n_1$ and $n_2$. \textbf{b,} Varying the polarization measurement on photon A allows us to traverse the topological landscape of the spin–skyrmion entangled state, revealing a quantum multiskyrmion composed of quasiparticle-like distributions at the equator of the sphere, and a higher order skyrmion at the poles (represented in the polarization vector plots where $\psi=\frac{1}{2}\mathrm{tan}^{-1}(S_2/S_1)$). The emergence of the different topological structures gives rise to a topological switch, yielding a skyrmion number of $n_1$ at the poles of the sphere, and $n_2$ at the equator.}
    \label{fig:concept}
\end{figure*}

\section*{Results}

\noindent \noindent \textbf{Non-local topological control.} To enable heralded non-local control of local topology, we begin with the quantum state  

\begin{equation}
    |\Psi\rangle_{AB} = |R\rangle_A|\phi_1\rangle_B +|L\rangle_A|\phi_2\rangle_B,
    \label{eqn:state}
\end{equation}

\noindent where the polarization of photon A is entangled with skyrmion states $|\phi_{1,2}\rangle$ ($n\neq0$) on photon B as illustrated graphically in Fig.~\ref{fig:concept}\textbf{a}. The skyrmion states are then defined as

\begin{eqnarray}
    |\phi_1\rangle_B = |R\rangle_B|\ell_1\rangle_B + |L\rangle_B|\ell_2\rangle_B, \\
    |\phi_2\rangle_B = |R\rangle_B|\ell_2\rangle_B + |L\rangle_B|\ell_3\rangle_B,
\end{eqnarray}

\noindent where $\ell_1$ and $\ell_2$ denote the OAM of $\ell_1 \hbar$ and $\ell_2 \hbar$ per photon respectively, with $|\ell_3| > |\ell_2| $, $|\ell_2| > |\ell_1| $, and $\ell_3 - \ell_2 = \ell_2 - \ell_1$ and \textit{R}, \textit{L} are the orthogonal right- and left-circular polarization states, respectively. Here, a skyrmionic topology emerges from the non-separability between the single photon's internal spin (polarization) and OAM degrees of freedom (DoFs). However, when measured on its own, the topology of photon B is, in fact, ‘unknown’ as photon B is described by a mixture of skyrmion states. To uncover the topology it carries, one must make a measurement on the polarization state of photon A. The topology of the single photon can thus be treated as a non-separable property of the state, manifesting only through a non-local measurement on its entangled partner. By deliberately controlling the polarization projection on photon A, we can remotely control and switch the skyrmion topology heralded on photon B as illustrated in Fig.~\ref{fig:concept}\textbf{a}. With this control, photon B can be made to carry either topological number $n_1$ or $n_2$, each corresponding to a distinct and uniquely structured skyrmion state. Our quantum entangled state thus realizes a non-local quantum dial for topological structure. 

\begin{figure*}[!htbp]
    \centering
    \includegraphics[width=1\linewidth]{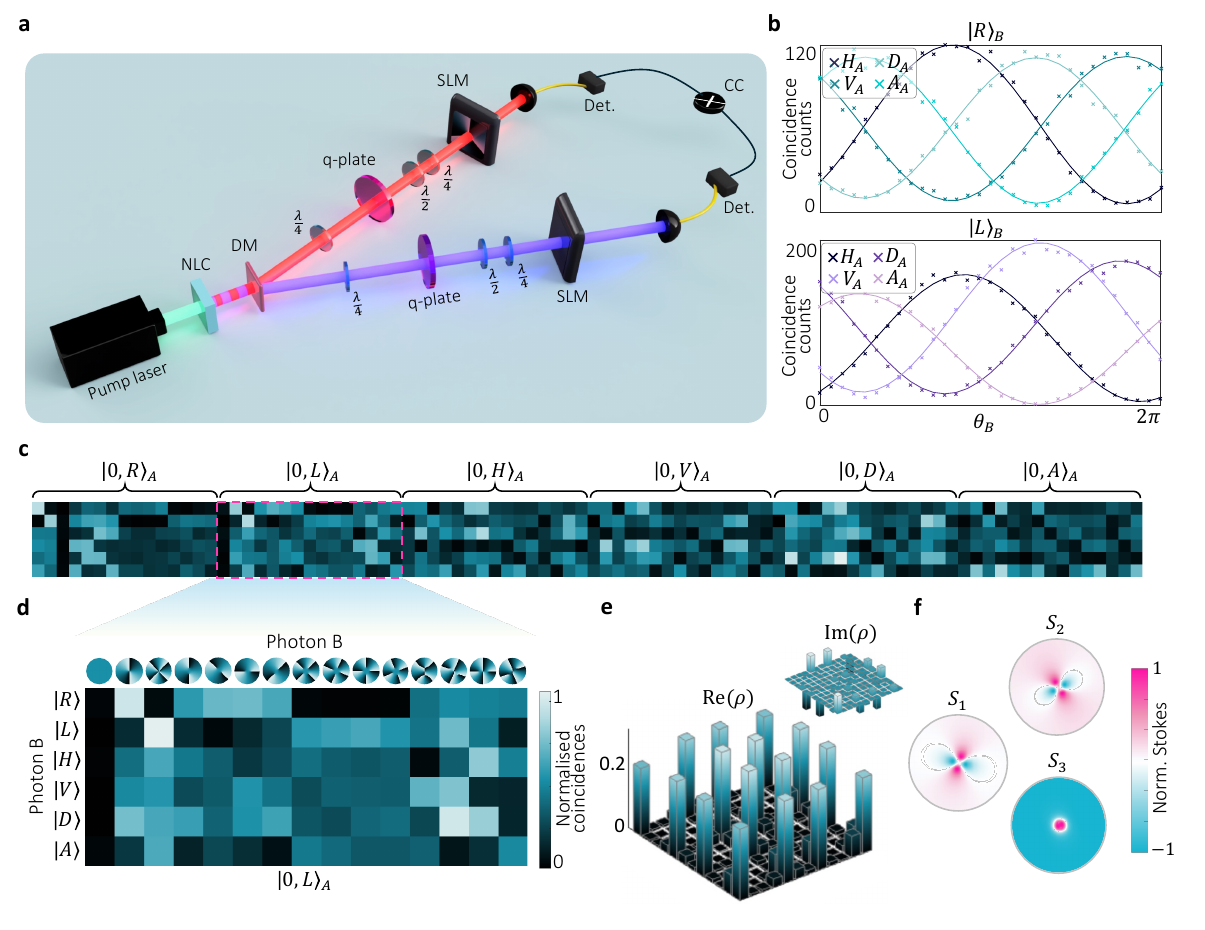}
    \caption{\textbf{Experimental generation of spin-skyrmion entangled states.} \textbf{a,} Schematic diagram of the non-degenerate quantum experiment used to generate and detect spin-skyrmion entangled states. A non-linear crystal (NLC) is used to produce dual-wavelength SPDC at $\lambda_A$ = 1550 nm and $\lambda_B$ = 810 nm. A dichroic mirror (DM) separates the wavelengths into two arms, with a q-plate and polarization optics (quarter- and half-wave plates) in each path. Both paths are directed to SLMs and thereafter coupled into SMFs connected to single photon detectors. Coincidences are recorded by a coincidence counter (CC) within a 0.5 ns window. To confirm the non-local nature of the states we generate, we demonstrate a \textbf{b,} violation of the Bell inequality for an example state where $\ell_1 = 0, \ell_2 = -2$ and $ \ell_3 = -4$. \textbf{c,} The measurement matrix for the QST performed on the example state, constructed by 6 polarization measurements on photon A, and 15 spatial measurements coupled with 6 polarization measurements on photon B, shown in the zoomed-in insets in \textbf{d}. \textbf{e,} Corresponding experimentally reconstructed density matrix extracted from the full QST with real and imaginary components. \textbf{f,} Stokes measurements retrieved directly from the tomography data.}
    \label{fig:experiment}
\end{figure*}

\noindent In practice, a projection onto the basis polarization states ($|R\rangle, |L\rangle$) for photon A heralds a particular quantum skyrmion state on photon B, either $|\phi_1\rangle$ or $|\phi_2\rangle$. These states are topologically equivalent, characterised by the same skyrmion number since $\ell_3 - \ell_2 = \ell_2 - \ell_1$. However, a more generalized polarization projection on photon A, $|P\rangle\langle P|$, where $|P\rangle = \text{cos}(\theta/2)|R\rangle + \text{sin}(\theta/2) e^{i\alpha}|L\rangle$ and $\theta \in [0, \pi], \alpha \in [0, 2\pi]$, results in the new state

\begin{equation}
    |\Psi\rangle_{B|A} = \text{cos}(\theta/2)|\phi_1\rangle_B +\text{sin}(\theta/2) e^{-i\alpha}|\phi_2\rangle_B.
    \label{eq:PhotonAProj}
\end{equation}

\noindent By adjusting the relative amplitude (via $\theta$) and phase ($\alpha$) tuning angles, the polarization projection on photon A is varied, collapsing the state of photon B into either a single skyrmion state or superpositions thereof. As depicted in Fig.\textbf{~\ref{fig:concept}b}, the superposition states define a unique topology, revealing a particular topological number at the equator of the sphere, while the single skyrmion states at the poles contribute an entirely different number, transforming the sphere into a map of topology --- a topological Bloch sphere --- where the basis vectors are themselves topological states.

\noindent Traversing this topological landscape and visualising the polarization fields at different points on the sphere reveals the underlying structures that give rise to the topological transition of photon B. For an example state with $\ell_{1,2,3} = 0,-2,-4$, these structures are visualised in Fig.~\ref{fig:concept}\textbf{b}. At the north pole of the sphere, the polarization texture represents a higher-order skyrmionic structure, yielding a skyrmion number of $n_1 = -2$. However, for the superposition states off the poles, a multiskyrmion structure emerges, revealing quasiparticle-like distributions, where the quasiparticles possess individual skyrmion numbers, as shown in Fig.~\ref{fig:concept}\textbf{b}. The total skyrmion number is found by a sum of the individual localized numbers $m_j$, as $n = \sum_jm_j$, giving a wrapping number of $n_2=-4$ for the superposition states in our example. The skyrmion number for photon B is thus parametrized by the non-local measurement angles ($\theta, \alpha$) of photon A,



\begin{equation}
    n(\theta, \alpha) = \frac{1}{4\pi}\int_{\mathcal{R}^2} \epsilon_{ijk} S_i(\theta,\alpha) \frac{\partial S_j(\theta,\alpha)}{\partial x} \frac{\partial S_k(\theta,\alpha)}{\partial y} dx dy,
\end{equation}


\noindent where $\epsilon_{ijk}$ is the Levi-Civita symbol and $S_{i,j,k} = S_{1,2,3}$ are the local normalised spatially varying Stokes parameters where $\sum^3_{i=1}S^2_i = 1$, ensuring a mapping to the unit sphere. 





\noindent To verify the non-local topological control, we use the experimental setup depicted in Fig.~\ref{fig:experiment}\textbf{a}. Dual wavelength photon pairs entangled in the spatial degree of freedom are generated in a Type 0, non-linear crystal of length $l = 5$ mm, via spontaneous parametric downconversion (SPDC). The photon pairs are separated using a dichroic mirror, reflecting photons of wavelength $\lambda_A = 1550$ nm and transmitting photons of wavelength $\lambda_B = 810$ nm. In each arm, the crystal plane was imaged to have a beam diameter of $2w$ = 0.45 mm on an electrically-tunable q-plate \cite{slussarenko2011tunable}. The q-plates have azimuthally-varying optical axes in the transverse plane with a topological charge $q$ which imparts $-2q\hbar$ ($2q\hbar$) OAM per photon for right (left) circularly polarized light at full conversion. With half-tuning, the transformation $\hat{U}_{QP}\ket{R,\ell} \rightarrow \frac{1}{\sqrt{2}}(\ket{R,\ell} + \ket{L,\ell-2q})$ was performed on each photon, where $\hat{U}_{QP}$ represents the q-plate operator at 50\% conversion efficiency. This transforms our initial OAM-entangled state to a spin-skyrmion coupled state, with details of the transformation given in the SI.  
The plane of each q-plate was then imaged onto a spatial light modulator (SLM). To perform the necessary projections and state tomography, we employed two waveplates and the SLM. Polarization projections were made on photon A while projecting onto the spatial mode $\ell_A = 0$, heralding a different skyrmion state on photon B, where both polarization and OAM DoFs were measured locally. A combination of a quarter-wave plate, half-wave plate, and linear polarizer was used to project onto various polarization states. Due to the polarization sensitivity of SLMs, the light emerging from the SLM was horizontally polarized, allowing the SLM to serve as a filter for the polarization projections. The half-wave plate was rotated to access linear polarization states, and the quarter-wave plate enabled projection onto circular polarization states. After the SLMs, the photons were coupled to single-mode optical fibers (SMFs) connected to avalanche photo-diodes (APDs) for photon detection. The photon counting device (CC) measured the coincidences within a 0.5 ns detection window. 

\noindent To confirm the non-local nature of the state with $\ell_{1,2,3}=0, -2, -4$, in this experiment (created using q-plates with $q = 1$), we performed measurements to demonstrate a violation of the Clauser–Horne–Shimony–Holt (CHSH) Bell inequality. Joint projective measurements were carried out between photons A and B of our quantum state by selecting the polarization settings $\{H,V,D,A\}$ on photon A and rotating the angle ($\theta_B$) of the spatial analyser $\ket{\theta_B}=\ket{0} + e^{-i\theta_B}\ket{-2}$ for each setting in the $\ket{R}$ subspace of photon B as well as $\ket{\theta_B}=\ket{-2} + e^{-i\theta_B}\ket{-4}$ in the $\ket{L}$ subspace. Here, H,V,D and A respectively mean horizontal, vertical, diagonal and antidiagonal polarization states. Our results demonstrate a violation of the Bell inequality with a Bell parameter of $S= 2.53$ for the heralded state $\ket{R}_A\ket{0}_B + \ket{L}_A\ket{-2}_B$ and $S= 2.47$ for state $\ket{R}_A\ket{-2}_B + \ket{L}_A\ket{-4}_B$, with Bell curves shown in Fig.~\ref{fig:experiment}\textbf{b}.

\begin{figure*}[t]
    \centering
    \includegraphics[width=1\linewidth]{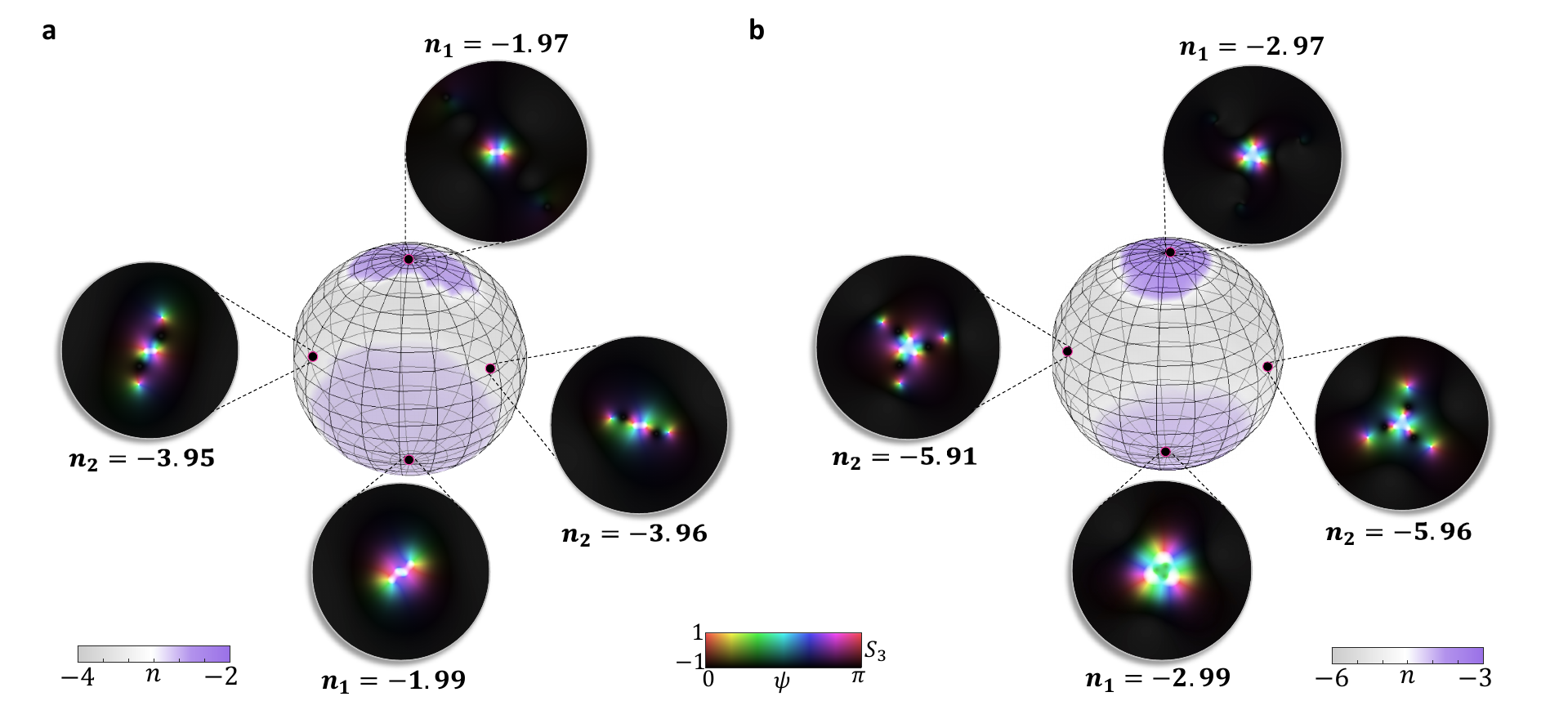}
    \caption{\textbf{Experimental verification of non-locally controlled topology.} A switch in the topological skyrmion number of photon B is realised through non-local polarization measurements on photon A. The spin textures are illustrated at different points on the sphere for a state with \textbf{a,} $\ell_1 = 0, \ell_2 = -2$ and $ \ell_3 = -4$, yielding a switching of the skyrmion number from $n_1=-2$ at the poles to $n_2=-4$ at the equator and \textbf{b,} $\ell_1 = 0, \ell_2 = -3$ and $ \ell_3 = -6$, switching between the two distinct skyrmion numbers, $n_1=-3$ and $n_2=-6$.}
    \label{fig:SwitchResults}
\end{figure*}

\noindent We demonstrate non-local switching of the generated local topology by performing a full quantum state tomography (QST) on the entangled photon state, following a similar procedure to Ref. \cite{d2016entangled}. For our state, this corresponds to six polarization projections on the heralding photon, and 15 spatial projections coupled with 6 polarization projections on photon B, resulting in an overcomplete QST with 540 entries shown in Fig.~\ref{fig:experiment}\textbf{c}. The density matrix (Fig.~\ref{fig:experiment}\textbf{e}) is then reconstructed with a fidelity $F = 0.93$ and purity $\gamma = 0.96$. Details of the density matrix construction from these measurements and of computing the corresponding entanglement witnesses is outlined in the SI. To analyse the skyrmionic structure of our state, the Stokes parameters (Fig.~\ref{fig:experiment}\textbf{f}) are extracted from the density matrix. 


\noindent Fig.~\ref{fig:SwitchResults} represents the experimental verification of the non-local topological switch for two example states, where $\ell_{1,2,3}=0, -2, -4$ (Fig.~\ref{fig:SwitchResults}\textbf{a}) and $\ell_{1,2,3}=0, -3, -6$ (Fig.~\ref{fig:SwitchResults}\textbf{b}). Here, the latter state was generated using q-plates with q=1.5. The spin textures are visualised at different points on the topological sphere in Fig.~\ref{fig:SwitchResults}, representing the skyrmionic structure of our state. In Fig.~\ref{fig:SwitchResults}\textbf{a}, a skyrmion number of $n_1\approx-2$ is observed at the poles of the sphere, whereas a multiskrymion structure is realized at the equator, yielding $n_2\approx-4$. The same behavior is observed for the state in Fig.~\ref{fig:SwitchResults}\textbf{b}, where the number of quasiparticles in the multiskyrmion structure now increases to three, producing a total skyrmion number of $n_2\approx-6$ at the equator, whereas the higher-order skyrmion at the poles has a wrapping number of $n_1 \approx-3$. Notice that the portion of the spheres corresponding to the basis skyrmion states extends beyond just the polar regions, due to experimental crosstalk in the OAM subspaces measured. This spread is more confined for the $\ell_{1,2,3}=0,-3,-6$ subspace where the wider separation in OAM values minimises crosstalk effects. These results experimentally confirm the non-local influence of photon A on the local topological structure of photon B, realizing a non-local skyrmion number switch. Our experimental results also confirm the expected multiskyrmion structure at the equator for both cases. While these results demonstrate a binary switch (between two different skyrmion numbers), we can extend this further by varying the type of heralded projection made on photon A, producing a ternary switch (between three distinct topological numbers), and also easily extending our skyrmion structures to higher-order skyrmions and multiskyrmions with more quasiparticles. This extended case is described in detail in the SI, accompanied by numerical simulations that confirm the ternary switch. \\

\begin{figure*}[!ht]
    \centering
    \includegraphics[width=1\linewidth]{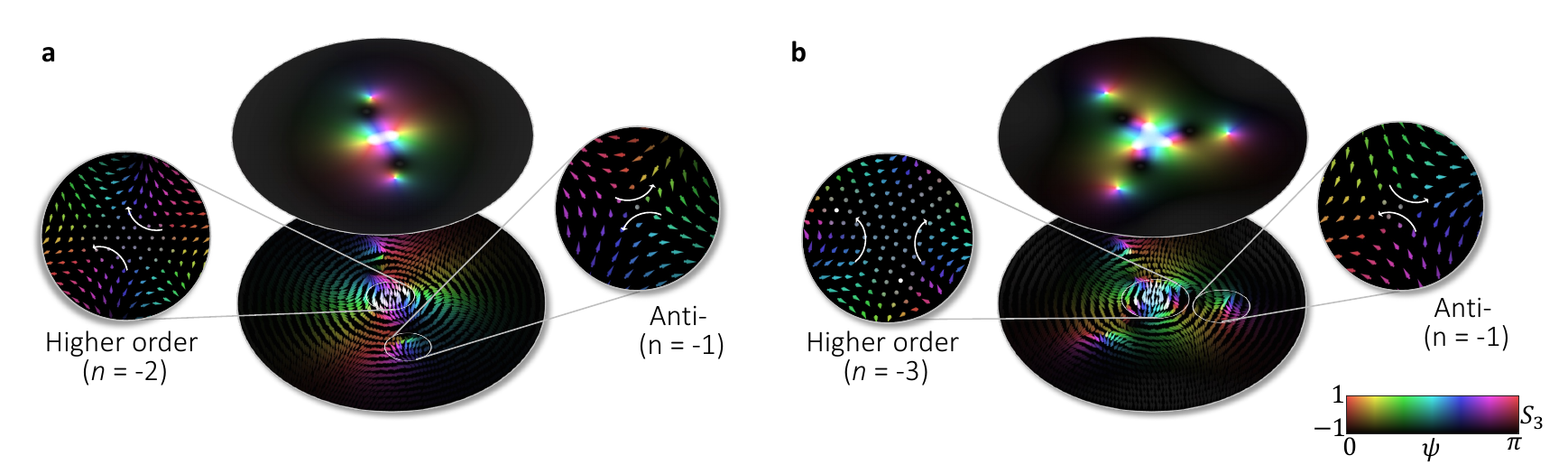}
    \caption{\textbf{Experimental quantum multiskyrmions.} Spin textures of the different flavours of quantum multiskyrmions realised, with zoom-in insets revealing the embedded local polarization texture of the quasiparticles in the structure for a state with \textbf{a,} $\ell_1 = 0, \ell_2 = -2$ and $ \ell_3 = -4$, and \textbf{b,} $\ell_1 = 0, \ell_2 = -3$ and $ \ell_3 = -6$.}
    \label{fig:Multiskyrmionresults}
\end{figure*}

\noindent \textbf{Quantum multiskyrmion.} Thus far we have studied how the topology of photon B is affected by a measurement on photon A, whilst remaining agnostic to photon B's ``local" polarization texture. We now instead look directly at the quantum multiskyrmion structure revealed via the non-local control. Fig.~\ref{fig:Multiskyrmionresults} shows a visualisation of the spin texture of the two flavours of quantum multiskyrmions created in our experiment. The polarization texture of the multiskyrmion reveals the distinct local textures embedded within the broader field. Zooming into the individual skyrmions shows the hyperbolic polarization texture characteristic of the anti-skyrmion for the quasiparticle in each case, and the expected signature with increased vorticity is seen for the central higher-order skyrmion \cite{shen2024optical}. Note that our scheme is not restricted to the aforementioned skyrmion and multiskyrmion textures, but can be readily extended to a broader class of textures even within the multiskyrmion structure. This is achieved by simply modifying the input polarization to the q-plates, which enables tuning from negative to positive skyrmion numbers and generates distinct localized skyrmion textures, such as Néel, Bloch, and intermediate-type, within a single multiskyrmion structure. These variations are illustrated in the simulations presented in the SI for the ternary switch case. \\

\begin{figure*}[!ht]
    \centering
    \includegraphics[width=1\linewidth]{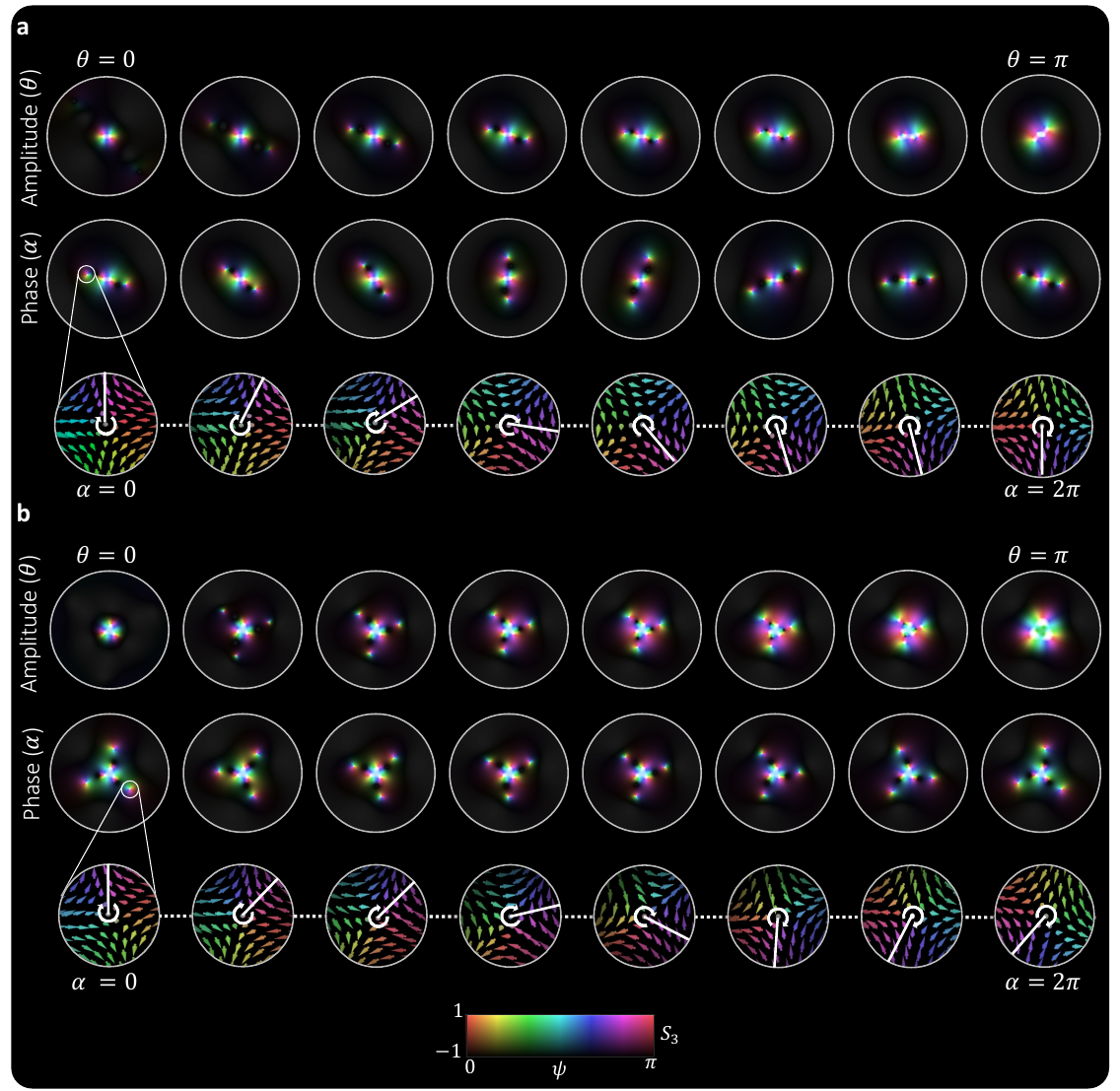}
    \caption{\textbf{Quasiparticle dynamics.} Non-locally derived polarization textures where the localized quasiparticle skyrmion vector distributions shift in and out radially when varying $\theta$ for $\alpha$ fixed at 3.77 rad from left to right, or exhibit an orbital and spin motion when varying $\alpha$ for $\theta$ fixed at $1.26$ rad, for the state with \textbf{a,} $\ell_1 = 0, \ell_2 = -2$ and $ \ell_3 = -4$, and for the state with \textbf{b,} $\ell_1 = 0, \ell_2 = -3$ and $ \ell_3 = -6$. In \textbf{a,} the amplitude ($\theta$) varies from left to right as $\theta=[0, 0.31, 0.63, 0.94, 1.26, 1.57, 2.20, 2.5, 3.14]$ rad and the phase ($\alpha$) as $\alpha=[0, 0.62, 1.26, 1.88, 3.14, 3.77, 5.03, 6.28]$ rad and in \textbf{b,} $\theta=[0, 0.94, 1.57, 1.88, 2.20, 2.83, 3.14]$ rad and $\alpha=[0, 0.62, 1.26, 1.88, 2.51, 3.14, 3.77, 4.40, 6.28]$ rad.}
    \label{fig:quasiparticle}
\end{figure*}

\noindent \textbf{Non-locally derived quasiparticle dynamics.} Having verified the existence of a multiskyrmion structure with localized quasiparticles, we are now interested in studying the dynamics of the polarization texture of photon B as we incrementally change the projection angles on photon A. In Fig.~\ref{fig:quasiparticle} the quasiparticle distributions of photon B's skyrmion state are shown as the amplitude, $\theta$, and phase, $\alpha$, are varied continuously. Results for the state with OAM values $\ell_{1,2,3}= 0,-2,-4$ and $\ell_{1,2,3}= 0,-3,-6$ are shown in \textbf{a} and \textbf{b}, respectively. A radial motion of the quasiparticles is observed when varying the amplitude, with the quasiparticles approaching the central distribution from infinity as $\theta$ increases from $0$ to $\pi$. This behaviour is mimicked in both sets of results with the only difference being in the number of quasiparticles experiencing these dynamics: $2$ and $3$ quasiparticles for \textbf{a} and \textbf{b}, respectively. For $\theta$ angles close to $\pi$, the quasiparticles start to merge with the central distribution inhibiting the ability to detect them as separate localized distributions. Varying the phase ($\alpha$) results in the orbit of the quasiparticles about the central distribution with increasing $\alpha$ from $0$ to $2\pi$. In addition to the orbiting behaviour, the quasiparticles also exhibit a local spin which is represented in the insets showing the evolution of the selected quasiparticle as the rotation of the phase profile $\psi$. This is induced by a local precession of each  Stokes vector about the $S_3$ axis. Whilst this behaviour is also mimicked in both sets of results, the symmetry of the quasiparticle distribution plays a role in the total spin and orbit angles performed by each quasiparticle within the distribution. Under a full scan of phase projection angles $\alpha$, the quasiparticles in \textbf{a} perform an orbit and spin of $\pi$ and in \textbf{b} exhibit an orbit of $\frac{2\pi}{3}$ and spin of $\frac{4\pi}{3}$, which is consistent with the 2- and 3-fold symmetries of their respective distributions. \\

\begin{figure}[!ht]
     \centering
     \includegraphics[width=1\linewidth]{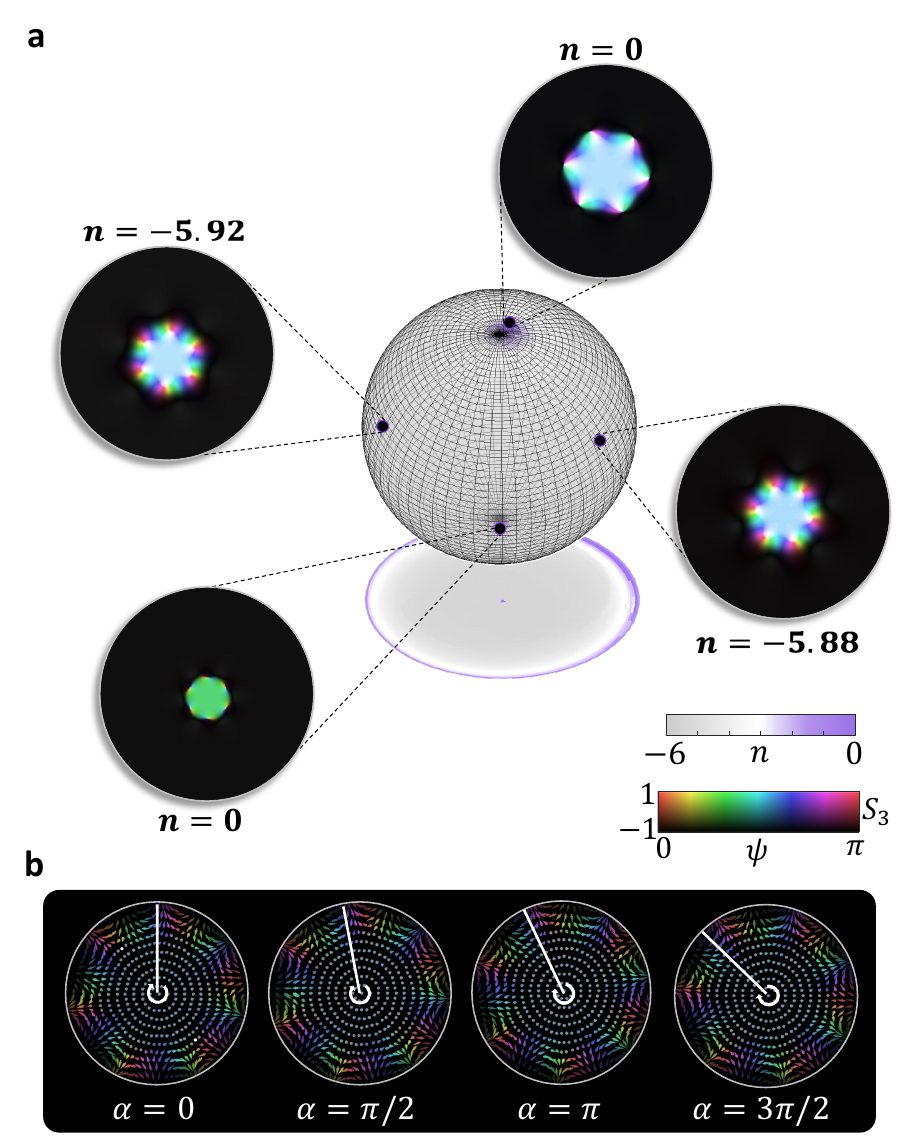}
     \caption{\textbf{Topology of the embedded GHZ-like states.} \textbf{a} The topological landscape of the embedded GHZ-like state reveals a trivial topology ($n=0$) at the poles of the sphere, transforming into states with skyrmion number $n=-6$ at the equator which correspond to non-separable states. \textbf{b} Spin textures of Bell states for an equatorial state.}
     \label{fig:GHZ}
 \end{figure}

\noindent \textbf{Embedded GHZ state and topological structure.}
Next, we show that a GHZ-like state can be isolated from our tripartite state, exhibiting an evolving spin texture emerging from the defining signature of GHZ states: a superposition projection on one particle of a GHZ state leaves the remaining two particles in an entangled Bell state. Here our skyrmion states inspire a new visualization of this effect; the Bell states are essentially skyrmions with transforming spin textures as illustrated in Fig.~\ref{fig:GHZ}\textbf{a}, which can be smoothly deformed into quasiparticle systems with more complex dynamics.

To see this, we remind the reader of the previous section, where a superposition measurement on photon A collapsed the state of photon B into a non-separable state exhibiting correlations between its internal spatial and polarization DoFs, producing complex orbital and spin dynamics. 
This is indeed reminiscent of the aforementioned tripartite GHZ state which collapses to an entangled Bell state upon a similar projective measurement. Specifically, by constraining photon B to the spatial Hilbert space spanned by the OAM states $|\ell_1\rangle$ and $|\ell_3\rangle$ we can extract the following GHZ-like state
 \begin{equation}
    |\Psi_{\text{GHZ}}\rangle_{AB} = \frac{1}{\sqrt{2}}\left(|R\rangle_A|R\rangle_B|\ell_1\rangle_B+ |L\rangle_A|L\rangle_B|\ell_3\rangle_B \right). 
 \end{equation}
 As before, performing polarization measurements on photon A then reveals the topological sphere and behaviour of the embedded GHZ state as shown in Fig.~\ref{fig:GHZ} for $\ell_{1,2,3} = 0,-3,-6$ with \textbf{a} showing the topological space and \textbf{b} confirming the texture for the 4 superposition measurements ($\alpha=0,\pi/2,\pi,3\pi/2$). (Further details are provided in the SI). Projections onto the basis states collapse our GHZ-like state into the separable states $|R\rangle_B|0\rangle_B$ and $|L\rangle_B|-6\rangle_B$ each carrying a trivial topology ($n=0$) depicted at the poles of the topological sphere shown in Fig.~\ref{fig:GHZ}\textbf{a}. However, projections onto the superposition states defined as $|\alpha\rangle_A = |R\rangle_A + e^{-i\alpha}|L\rangle_A$ collapse our GHZ-like state into non-separable states of the form $|R\rangle_B|0\rangle + e^{-i\alpha}|L\rangle_B|-6\rangle$ characterized by a skyrmion number $n=-6$ depicted at the equator of the sphere shown in Fig.\ref{fig:GHZ}\textbf{a}. The four Bell states are then obtained by projecting onto states
 with $\alpha=\{0,\pi,\pi/2,3\pi/2\}$ with their corresponding spin textures shown in \textbf{b}. Each Bell state corresponds to a different topological texture of a single central quasiparticle. Since the Bell states and the states considered in Eq.\ref{eq:PhotonAProj} (for the same polarization measurements) are characterized by identical skyrmion numbers, they can be smoothly deformed into one another. Here the smooth deformation is driven by interfering the GHZ state with a reference state $|\Psi_\text{ref}\rangle_{AB} = \left(|R\rangle_A|L\rangle_B + |L\rangle_A|R\rangle_B\right)|\ell_2\rangle_B$. Therefore, the Bell state signature of the embedded GHZ state manifests physically as diverse quasiparticle dynamics derived from the localized spin-texture of photon B given a superposition measurement on photon A.\\ 

 \noindent In this way, we establish an intrinsic link between the entanglement dynamics of GHZ states and the evolution of their underlying topological structure. 

\section*{Discussion and Conclusion}
\noindent Whilst the initial driving force behind the search for skyrmionic topological features in optical systems has been spurred on by their promise as robust information carriers within magnetic-spin systems, the field has since matured into demonstrations of their resilience across various optical systems such as classical vector beams \cite{wang2024topological, guo2026topological}, quantum non-local entangled states \cite{ornelas2025topological, ornelas2024non, guo2026topological} and single photon spin-orbit coupled states \cite{wang2025topological}. This has further fueled investigations into classifying more complex and diverse optical systems by their topological characteristics. In particular, quantum entangled states present a unique set of opportunities where the topological classification of high-dimensional entangled states and multi-particle states with diverse entanglement structures are almost completely unexplored. This work extends the notion of the topological classification of quantum states to a tripartite entangled state through the introduction of a ``topological Bloch sphere" highlighting the existence of multiple topological transition points, and connecting this work to recent discoveries which reveal the topological spectrum of high-dimensional entangled states \cite{de2025revealing}. Our skyrmion state preparation protocol builds upon remotely controlled topological systems \cite{corona2024generation}, paving the way towards the transmission of this robust resource, granting one party access to topological structures that would otherwise be inaccessible due to limited experimental capabilities. By simply sharing one photon with a resource-scarce partner, complex skyrmions with rich internal structure and dynamics, exceeding even dual topological switches (see SI), can be remotely transmitted for use as non-classical resources in quantum applications, where robustness to noise and high-dimensional structures are advantageous. Furthermore, our work reveals an intrinsic link between entanglement features and skyrmion topologies, extending the notion of visualising entanglement \cite{moreau2019imaging, fickler2013real} by leveraging topological quasiparticles and thereby paving the way for the physical observation of multipartite state evolution. In this case, rather than remotely controlling the quasiparticle dynamics, one can use the evolving multiskyrmion structure of one photon as a probe to infer the effect of a perturbation on its entangled partner, opening exciting possibilities for quantum topological sensing. We therefore envision our work driving new research avenues focused on identifying and exploiting the complex topological structures of high-dimensional and multi-particle states.\\

\noindent In conclusion, we have presented a unique realization of quantum topology in the form of spin-skyrmion entangled states, realizing a remotely controlled topological switch between distinct topological states. Through this experimental verification, we emulate structures seen in magnetic systems, realizing the first quantum multiskyrmions reminiscent of magnetic biskyrmions \cite{gobel2019forming}. Our non-local control enables us to replicate the spin and orbital dynamics of the quasiparticles, advancing beyond static optical skyrmions to non-locally tunable quantum multiskyrmions with quasiparticle dynamics. The topological states we create contain embedded GHZ-like states with entanglement characteristics manifesting physically as evolving topological quasiparticle structures. 
While our approach is compatible with degenerate quantum-entangled systems, our experimental realization in a non-degenerate entanglement setup extends the versatility of the approach to dual-wavelength systems, allowing topological control and detection to be implemented across configurations benefiting from interactions at different wavelengths. For instance, non-linear quantum transport \cite{sephton2023quantum} of the topological states, remote controlled transfer of the skyrmionic topology to matter \cite{mitra2025topological}, or topological interaction at advantageous wavelengths, while controlled with technologically accessible ones \cite{lu2019chip,rielander2014quantum, clausen2014source}. We expect these findings to open new avenues for harnessing and dynamically manipulating quantum skyrmions and multiskyrmions for quantum information encoding protocols and quantum topological sensing of evolving systems. \\

\section*{Acknowledgements}
F.N. and P.O acknowledge funding from the Council for Scientific and Industrial Research under the HCD-IBS scholarship scheme. M.K acknowdedges funding from the National Research Foundation (NRF). A.F and I.N acknowledge funding from the South African Quantum Technology Initiative (SA QuTI). V.D.A. and B.S acknowledge the funding from the Italian Ministry of Research (MUR) through the PRIN 2022 project QNoRM and the PNRR project PE0000023-NQSTI. B.P. also acknowledges financial support from the PRIN MUR project, CUP E53D23002520006.

\section*{Author contributions}
F.N and B.S. designed, built and performed the experiment and analysed the data. B.P. designed the q-plates used in the experiment. P.O and M.K contributed code for analysing the data. P.O conceived of the idea. A.F and V.D.A supervised the project. All authors contributed to the writing of the manuscript.

\section*{Competing interests}
The authors declare no competing interests.

\section*{Data availability}
Data supporting this study are available from the authors upon reasonable request.

\bibliographystyle{unsrt}
\bibliography{references.bib}

\pagebreak
\widetext

\begin{center}
    \textbf{\large Supplementary information: Remote engineering of particle-like topologies to visualise entanglement dynamics}
\end{center}

\section{Non-local topological control with arbitrary $\ell_A$ projections on photon A}

\noindent Here we consider the OAM-entangled SPDC state following the insertion of identical half-tuned q-plates with charge $q$ in both arms, and examine the effect of arbitrary OAM projections on the heralding photon. We begin with a general OAM entangled state (assuming R polarization)

\begin{equation}
   \ket{\Psi} \!= \!\sum_{\ell \in \mathbb{Z}} c_{\ell,-\ell} \ket{\ell}_A \ket{R}_A \!\ket{-\ell}_B \ket{\rm R}_B,
\end{equation}
\noindent where $c_{\ell,-\ell}$ is the the two-photon probability amplitude. The half-tuned q-plate transformation, $\hat{U}_{QP}\ket{R,\ell} \rightarrow \frac{1}{\sqrt{2}}(\ket{R,\ell} + \ket{L,\ell-2q})$ with $\hat{U}_{QP}$ representing the q-plate operator at 50\% conversion efficiency, is applied on both photons producing

\begin{equation}
    \ket{\Psi} = \frac{1}{2}\sum_{\ell \in \mathbb{Z}}c_{\ell,-\ell} \left\{ \ket{\ell}_A\ket{R}_A \left( \ket{R}_B\ket{-\ell}_B + \ket{L}_B\ket{-\ell-2q}_B \right) + \ket{\ell-2q}_A\ket{L}_A \left( \ket{R}_B\ket{-\ell}_B + \ket{L}_B\ket{-\ell-2q}_B \right) \right\},
\end{equation}

\begin{figure*}[!ht]
    \centering
    \includegraphics[width=1\linewidth]{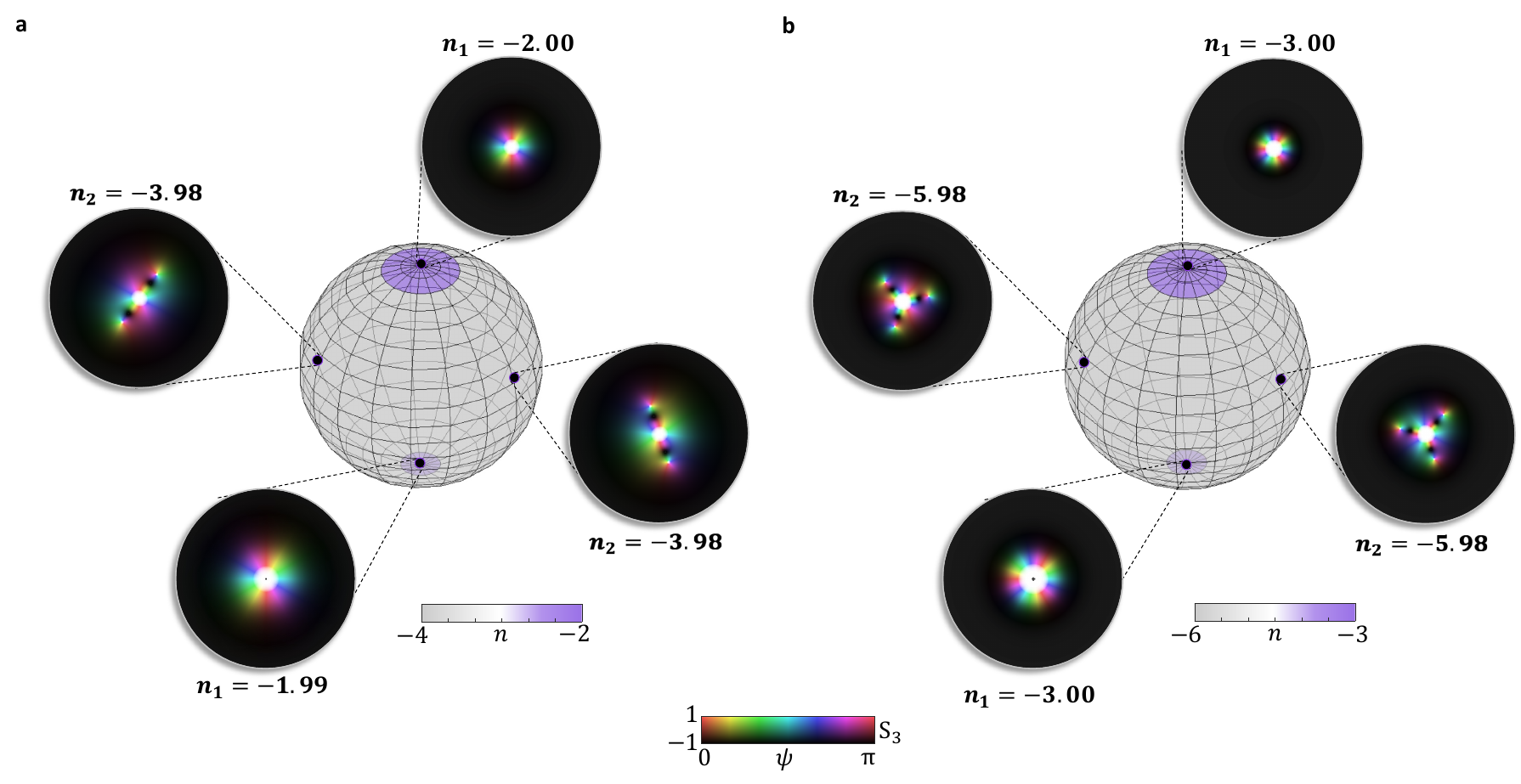}
    \caption{\textbf{Simulated binary topology switch results.} A binary switch in the topological skyrmion number of photon B is realised, with the spin textures illustrated at different points on the sphere for a state with \textbf{a,} $\ell_{1,2,3} = 0,-2,-4$, yielding a switching of the skyrmion number from $n_1=-2$ at the poles to $n_2=-4$ across the remaining sphere and \textbf{b,} $\ell_{1,2,3} = 0,-3,-6$, switching between the two distinct skyrmion numbers, $n_1=-3$ and $n_2=-6$.}
    \label{fig:simulationsforexp}
\end{figure*}

\noindent For a projection of photon A onto the fundamental Gaussian mode, $\ell_A=0$, we create the spin-skyrmion entangled states presented in the main text with simulated results for the cases with $\ell_{1,2,3} = 0,-2,-4$ and  $\ell_{1,2,3} = 0,-3,-6$ represented in Fig.~\ref{fig:simulationsforexp}, showing excellent agreement with the binary topological switch seen in the experimental results. The polarization distribution plots shown here correspond to different regions of the sphere for the superposition states, while still yielding a total skyrmion number of $n\approx-4$ and $n\approx-6$ for these states. Notably, the distribution on the sphere corresponding to $n=-2$ and $n=-3$ is more confined to the polar regions compared to the experimental results which are affected by crosstalk. Moreover, a finer sampling of the relative amplitude ($\theta$) and phase ($\phi$) angles would also produce a topological sphere with more localized polar regions. 

\noindent In the previous simulations and the experimental realization of this work, we considered the case where photon A was projected onto the mode $\ell_A = 0$. This projection, however, need not be confined to the fundamental mode in order to achieve a state in the form of Eq. 1 (main text), but can in principle be extended to any arbitrary value. Given a projection of photon A onto any arbitrary OAM mode, such as $\ell'_A$, the quantum state becomes 
 \begin{equation}
  \begin{split}
     \ket{\Psi'} &= \bra{\ell'_A}_A\ket{\Psi} \\&\propto \sum_{\ell \in \mathbb{Z}} \left\{ 
     \braket{\ell'_A|\ell}_A\ket{R}_A \left( \ket{R}_B\ket{-\ell}_B + \ket{L}_B\ket{-\ell-2q}_B \right) + \braket{\ell'_A|\ell-2q}_A\ket{L}_A \left( \ket{R}_B\ket{-\ell}_B + \ket{L}_B\ket{-\ell-2q}_B \right)
     \right\}\\
     &\propto \ket{R}_A \left( \ket{R}_B\ket{-\ell_A'}_B + \ket{L}_B\ket{-\ell_A'-2q}_B \right) + 
     \ket{L}_A \left( \ket{R}_B\ket{-\ell_A'-2q}_B + \ket{L}_B\ket{-\ell_A'-4q}_B \right)
      \end{split}
      \label{Eq:ArbEll}
 \end{equation}
\noindent with the orthogonality of OAM modes rendering terms $\braket{\ell'_A|\ell}$ and $\braket{\ell'_A|\ell-2q}$ non-zero only when $\ell = \ell'_A$ and $\ell = \ell'_A + 2q$, respectively. It follows that photon B is conditioned onto the spin-orbit coupled states, $\ket{\phi_1}_B = \ket{R}_B\ket{-\ell_A'}_B + \ket{L}_B\ket{-\ell_A'-2q}_B$ and $\ket{\phi_2}_B = \ket{R}_B\ket{-\ell_A'-2q}_B + \ket{L}_B\ket{-\ell_A'-4q}_B$, by projection onto a single OAM mode on photon A. This is in agreement with the general form of Eqs. 1-3 (main text) where $\ell_1 =-\ell_A'$, $\ell_2 =-\ell_A'-2q$ and $\ell_3 =-\ell_A'-4q$ is determined by the combined choice of spatial projection on photon A as well as the charge of the chosen q-plates. The available skyrmion topologies for switching is given by $n_1 = \ell_3 - \ell_2 = \ell_2 - \ell_1 = -2q$ and $n_2 = \ell_3 - \ell_1 = -4q$.  As a result, non-zero values of $\ell_A$ can be used to tailor the OAM subspaces in which the skyrmions are generated. However, it is strictly the charge of the q-plates which allows one to tailor the skyrmion numbers.   

\begin{figure}
    \centering
    \includegraphics[width=1\linewidth]{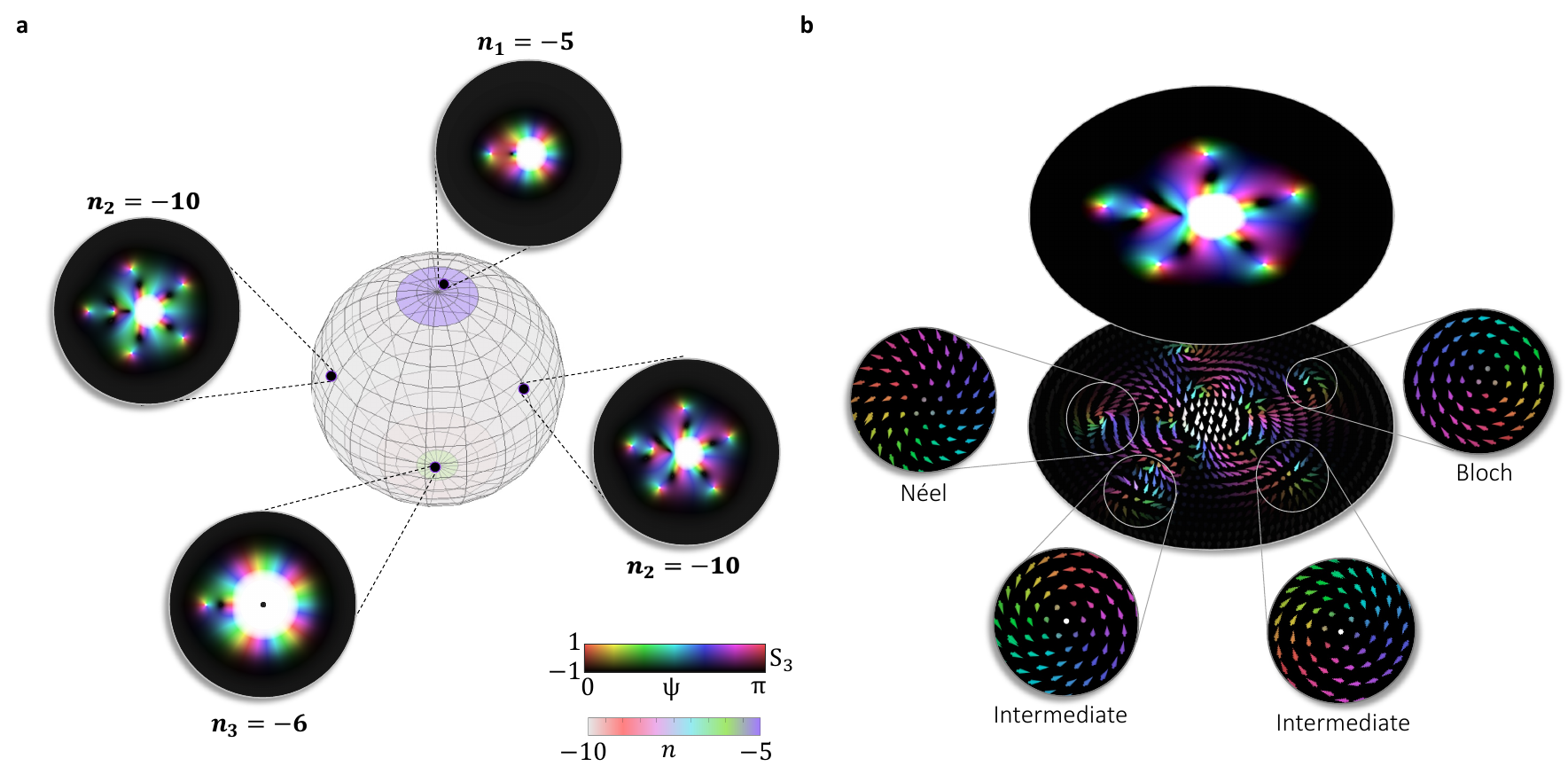}
    \caption{\textbf{Extended topological transitions.} \textbf{a,} A switch between three different topological numbers for photon B is realised for a simulated case. The spin textures are illustrated at different points on the sphere for a state with \textbf{a,} $\ell_{A_1} = 0$ and $  \ell_{A_2} = -1$, yielding a switching of the skyrmion number from $n_1=-5$ at the north pole to $n_2=-10$ at the equator and $n_3 = -6$ at the south pole. \textbf{b,} Simulated spin texture of a quantum multiskyrmion (with $n = 10$), with zoom-in insets revealing the embedded local polarization texture of the quasi-particles in the structure.}
    \label{fig:OAMsupProj}
\end{figure}

\noindent Interestingly, non-local control of extended topological transitions is possible by instead choosing to project photon A onto a superposition of OAM states. For instance, by projecting onto the two-mode superposition, $\ket{\ell_A} \rightarrow \frac{1}{\sqrt{2}}(\ket{\ell_{A_1}'}+\ket{\ell_{A_2}'})$ where $\ell_{A1}'$ and $\ell_{A2}'$ are different OAM values, the general quantum state becomes
 \begin{equation}
  \begin{split}
     \ket{\Psi'} &= \frac{1}{\sqrt{2}} \left[\bra{\ell_{A_1}'}+\bra{\ell_{A_2}'} \right] \ket{\Psi} \\
     &\propto \ket{R}_A \left( \ket{R}_B \left[\ket{-\ell_{A_1}'}_B +\ket{-\ell_{A_2}'}_B \right] + \ket{L}_B\left[\ket{-\ell_{A_1}'-2q}_B + \ket{-\ell_{A_2}'-2q}_B \right]\right) \space \\
     & \hspace{1pt} + \ket{L}_A \left( \ket{R}_B \left[\ket{-\ell_{A_1}'-2q}_B + \ket{-\ell_{A_2}'-2q}_B \right]+ \ket{L}_B\left[\ket{-\ell_{A_1}'-4q}_B + \ket{-\ell_{A_2}'-4q}_B\right] \right).
   \end{split}
 \end{equation}
\noindent Here, each polarization term of photon B, which was previously conditioned to hold a single OAM value, is now conditioned onto a superposition of the chosen OAMs; each of which undergo the same transformation of the q-plate as before (Eq. \ref{Eq:ArbEll}). To further explore the topological transitions, we choose $\ell_{A_1}=0$, $\ell_{A_2}=-1$ and $q = 2.5$ as an example, heralding the state
\begin{equation}
    \ket{\Psi'} \propto \ket{R}_A \left( \ket{R}_B \left[\ket{0}_B +\ket{1}_B \right] + \ket{L}_B\left[\ket{-5}_B + \ket{-4}_B \right]\right) 
    + \ket{L}_A \left( \ket{R}_B \left[\ket{-5}_B + \ket{-4}_B \right]+ \ket{L}_B\left[\ket{-10}_B + \ket{-9}_B\right] \right).
\end{equation}
\noindent The extended topological transitions that arise from this heralded state are represented in Fig.~\ref{fig:OAMsupProj}\textbf{a}, with the spin textures illustrated at different points on the topological landscape. Notably, this choice of projection introduces more quasi-particles into the multiskyrmion structures, now forming a multiskyrmion at the poles. The multiskrymion structure emerging at the poles of the sphere can be tied to the choice of OAM charge for $\ell_{A_2}$. (For example, if $\ell_{A_2} = -2$ instead, there would be two quasi particles in the multiskyrmion at the poles.) Additionally, a skyrmion number of $n_1=-5$ is observed at the north pole of the sphere, before transitioning to $n_2=-10$ for the equator states and $n_3=-6$ at the south pole.  
\noindent Furthermore, by adjusting the input polarization to the q-plates to left-circular polarization, one can create skyrmions and multiskyrmions with positive skyrmion numbers, enabling distinct skyrmion textures to coexist within a single multiskyrmion. The spin texture of such a multiskyrmion with $n_2=10$ is visualised in Fig.~\ref{fig:OAMsupProj}, revealing the distinct local textures embedded within the broader field. Zooming into the individual localized skyrmions reveals distinct textures for each, ranging from hedgehog and spiral polarization textures characteristic of the Néel and Bloch-type skyrmions respectively to intermediate textures. 

\section{Experimental results and data analysis}
\subsection{Density matrix reconstruction and Bell inequality violation }

\noindent To reconstruct the density matrix describing the quantum state of the bipartite system, we perform a quantum state tomography (QST) in the composite Hilbert space $\mathcal{H}=\mathcal{H}_{\mathrm{pol,A}} \otimes \mathcal{H}_{\mathrm{pol,B}} \otimes \mathcal{H}_{\mathrm{sp,B}}.$ Here photon~A is characterized solely by its polarization, while photon~B is described by both polarization and spatial degree of freedom. The density matrix $\rho$ can be expanded in a complete operator basis
\begin{equation}
    \rho =\frac{1}{d} \sum_{i,j,k} c_{ijk} (\sigma_{i} \otimes \sigma_{j} \otimes\lambda_{k})
\end{equation}
where $d=d_\mathrm{pol,A} \times d_\mathrm{pol,B} \times d_\mathrm{sp,B}$ is the total Hilbert space dimension, $\{\sigma_{i} \}$ and $\{\sigma_{j} \}$ denote Pauli operator basis for the polarization degrees of freedom for photons~A and~B, $\{\lambda_{k}\}$ are the generalised Gell-Mann operators spanning the spatial subspace of Photon~B, and $c_{ijk}$ are the state coefficients. 
We construct an informationally complete set of measurements from tensor 
products of projective measurements on each subsystem. The polarization measurements for both photons use eigenprojectors of the Pauli operators, 
whereas the spatial measurements on photon~B use eigenprojectors of the 
generalized Gell-Mann operators. Each measurement setting is represented by a 
 projector of the form $M_k = P_a^{(A,\mathrm{pol})}\otimes P_b^{(B,\mathrm{pol})}\otimes P_s^{(B,\mathrm{sp})},$ and the corresponding detection probability is given by Born rule $p_k = \mathrm{Tr}(\rho M_k)$. The reconstruction proceeds as a linear inversion problem. After vectorizing both the density matrix and the measurement operators, the forward model takes the 
form $\mathbf{p} = T\,\mathrm{vec}(\rho),
$ where $T$ is the measurement matrix assembled from all projectors. A physically valid estimate of $\rho$ is obtained by minimizing the least-squared norm of the residual \begin{equation}
  | \mathbf{p} - T\mathrm{vec}(\rho)|^{2}_{2}
\end{equation} between the experimentally measured probabilities and the Born-rule predictions. We enforce Hermiticity, positivity, and unit trace by parametrizing the density 
matrix in the form $\rho = \frac{L L^{\dagger}} {\mathrm{Tr}(LL^{\dagger})},$ where $L$ is a complex lower-triangular matrix with real exponential diagonal 
entries. This parametrization  guarantees positive semi-definiteness 
throughout the optimization.


\begin{figure*}
    \centering
    \includegraphics[width=1\linewidth]{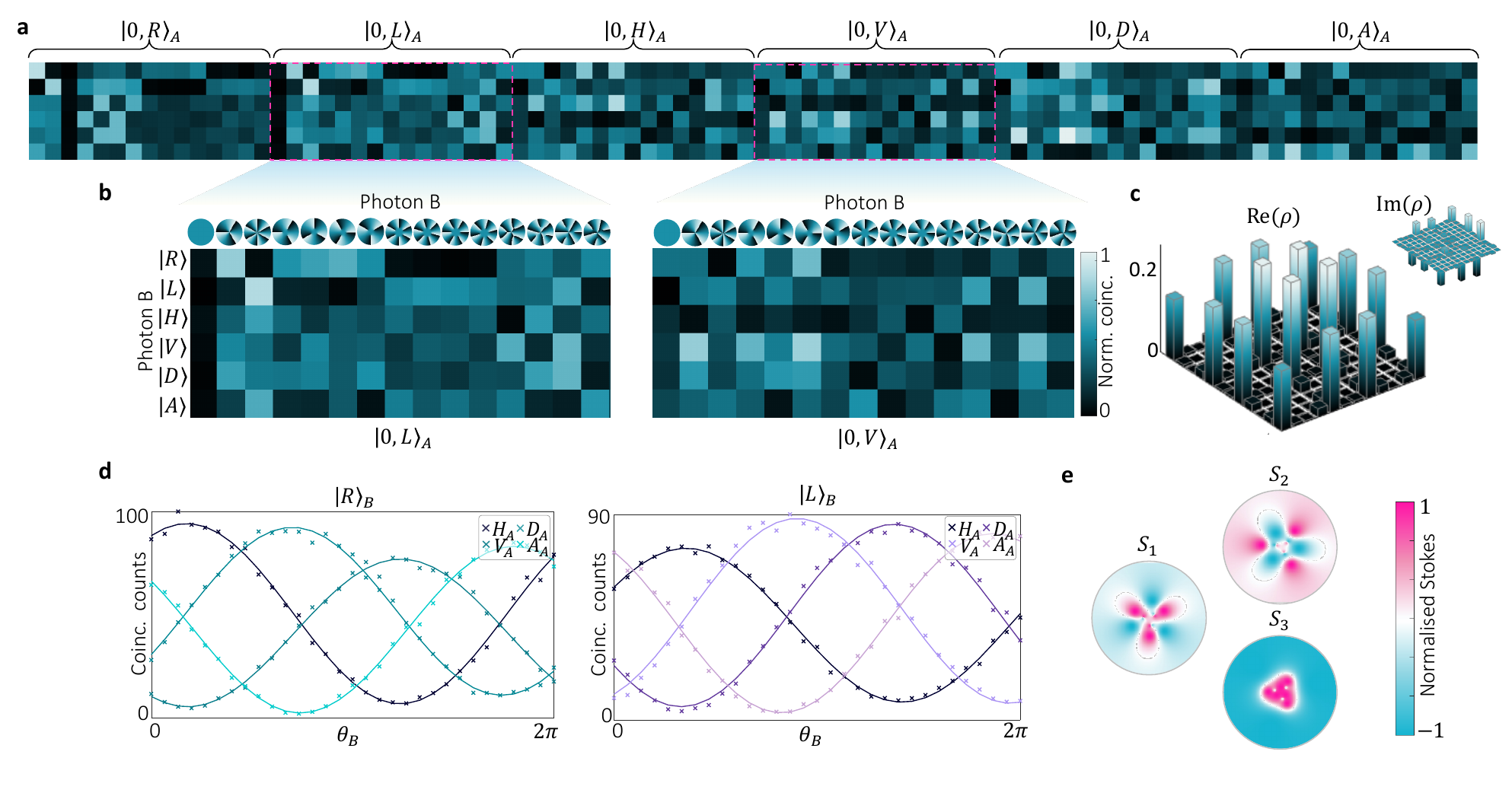}
    \caption{\textbf{Experimental characterisation results.} \textbf{a,} The measurement matrix for the QST performed on the example state with $\ell_1 = 0, \ell_2 = -3$ and $ \ell_3 = -6$, constructed by 6 polarization measurements on photon A, and 15 spatial measurements coupled with 6 polarization measurements on photon B, shown in the zoomed-in insets in \textbf{b}. \textbf{c,} Corresponding experimentally reconstructed density matrix extracted from the full QST with real and imaginary components. \textbf{d,} Violation of the Bell inequality. \textbf{e,} Stokes measurements retrieved directly from the tomography data.}
    \label{fig:QSTBellSupp}
\end{figure*}

\subsection{Entanglement Witnesses}

\subsubsection{Purity}
\noindent The purity, $\gamma$, of states provides an estimate of the total signal-to-noise ratio (SNR) within a given experiment \cite{zhu2021high}. Furthermore, the purity quantifies the degree of mixture of a state given by 

\begin{equation}
\gamma = \text{Tr} \left( \rho^2 \right) \geq \frac{1}{d}
\end{equation}

where $\gamma=1$ indicates a pure state and $\gamma = \frac{1}{d}$ indicates a maximally mixed state. For the states considered in this work $\gamma$ scales with the dimension of the spatial qudit such that $\gamma = \frac{1}{4d_{sp,B}}$.

\subsubsection{Fidelity}
\noindent The fidelity also served as an entanglement witness for our states, used to analytically compare our measured state, $\rho$ against the closest, pure maximally entangled state $\rho_T=|\Psi\rangle\langle\Psi|$
\begin{equation}
    F =\left( \text{Tr}  \left( \sqrt{  \sqrt{\rho_T}\rho \sqrt{\rho_T}  }  \right) \right)^2 ,
\end{equation}
 The fidelity is 0 if the states are orthogonal or 1 when they are identical up to a global phase. \\

\noindent Experimental results for the example state with $\ell_1 = 0, \ell_2 = -3$ and $ \ell_3 = -6$ (created using q-plates with $q = 1.5$) are shown in Fig.~\ref{fig:QSTBellSupp}. A full QST was performed on the entangled photon state represented by the tomography data in Fig.~\ref{fig:QSTBellSupp}\textbf{a} with zoomed in insets in \textbf{b}. The 
density matrix is then reconstructed as shown in Fig.~\ref{fig:QSTBellSupp}\textbf{c}, with fidelity, $F = 0.95$ and purity $\gamma = 0.97$. Furthermore, to verify the non-local nature of this example state, we performed measurements to demonstrate a violation of the Clauser–Horne–Shimony–Holt (CHSH) Bell inequality. Joint projective measurements were carried out between photons A and B of our quantum state by selecting the polarisation settings $\{H,V,D,A\}$ on photon A and rotating the angle ($\theta_B$) of the spatial analyser $\ket{\theta_B}=\ket{0} + e^{-i\theta_B}\ket{-3}$ for each setting in the $\ket{R}$ subspace of photon B as well as $\ket{\theta_B}=\ket{-3} + e^{-i\theta_B}\ket{-6}$ in the $\ket{L}$ subspace. Our results demonstrate a violation of the Bell inequality with a Bell parameter of $S= 2.43$ for the heralded state $\ket{R}_A\ket{0}_B + \ket{L}_A\ket{-3}_B$ and $S= 2.42$ for state $\ket{R}_A\ket{-3}_B + \ket{L}_A\ket{-6}_B$, with Bell curves shown in Fig.~\ref{fig:QSTBellSupp}\textbf{e}.

\subsection{Extraction of topological features}

\noindent The topological features of our tripartite state, $\rho$, are derived from the correlations between photon B's internal spin and spatial DoFs after performing a projective measurement on photon A's spin DoF. Therefore it is useful to map $\rho$ onto the spatially-varying polarization density matrix given  $\rho_{B|A}(\theta,\alpha,\vec{r}_B)$ which is given by  
\begin{equation}
\rho_{B|A}(\theta,\alpha,\vec{r}_B) = {}_A\langle \theta,\alpha|\, \mathbb{1}_{B,\text{pol}} \, {}_B\langle \vec{r}| \; \rho \; |\theta,\alpha\rangle_A \, \mathbb{1}_{B,\text{pol}} \, |\vec{r}\rangle_B,
\end{equation}
where $|\theta,\alpha \rangle_A = \text{cos}(\theta/2)|R\rangle_A + \text{sin}(\theta/2) e^{i\alpha}|L\rangle_A$ is an arbitrary polarization measurement on photon A with $\theta\in[0,\pi],\alpha\in[0,2\pi)$. In general we can describe our tripartite state as
\begin{equation}
\rho = \sum_{ijklmn}c_{ijklmn} \ket{P_i}_A \ket{P_j}_B \ket{\ell_k}_B\, {}_A\!\bra{P_l} {}_B\!\bra{P_m}\, {}_B\!\bra{\ell_n}
\end{equation}
where $\ket{P_{i,j,l,m}}\in\{\ket{R},\ket{L}\}$, $k,n\in\{0,1,2,...,d_{sp,B}\}$, and $c_{ijklmn}$ are the density matrix coefficients. The corresponding spatially-varying density matrix is then given by
\begin{equation}
\rho_{B|A}(\theta,\alpha,\vec{r}_B) = \sum_{jkmn} A_{jkmn}(\theta,\alpha) \; \text{LG}^{p=0}_{\ell_k}(\vec{r}_B) \; \left(\text{LG}^{p=0}_{\ell_n}(\vec{r}_B)\right)^* \; \ket{P_j}_B {}_B\!\bra{P_m} 
\end{equation}
where $\text{LG}^{p=0}_{\ell}(\vec{r}_B)$ are the Laguerre-Gaussian functions with radial order $p=0$, topological charge $\ell$, at $z=0$ that result from projecting the OAM modes onto space, i.e., ${}_B\bra{\vec{r}}\ell\rangle_B = \text{LG}^{p=0}_{\ell}(\vec{r}_B)$ and
$A_{jkmn}(\theta,\alpha)$ are the polarization projection coefficients that are dependent on the projection on photon A, given by
\begin{eqnarray}
A_{jkmn}(\theta,\alpha) = \sum_{jklm} &&(c_{0jklm0}\cos^2(\theta/2) + c_{0jklm1}\cos(\theta/2) \sin(\theta/2)e^{-i\alpha} + \nonumber\\
&&c_{1jklm0}\cos(\theta/2) \sin(\theta/2)e^{i\alpha} + c_{1jklm1}\sin^2(\theta/2)). 
\end{eqnarray}

Next, we extract the spatially-varying quantum Stokes vector, $\vec{S}_{B|A}(\theta,\alpha,\vec{r}_B)$ whose components are given by $S_{B|A,p} (\theta,\alpha,\vec{r}_B) = \text{Tr}\left( \sigma_p \,\rho_{B|A}(\theta,\alpha,\vec{r}_B)\right)$, where $\sigma_p$ are the usual Pauli-spin matrices with $\sigma_0=\mathbb{I}_2$. From this it is then straight-forward to verify that the quantum Stokes vector is given by
\begin{equation}
\vec{S}_{B|A}(\theta,\alpha,\vec{r}_B) = \begin{bmatrix}
    S_{B|A,0} (\theta,\alpha,\vec{r}_B) \\
    S_{B|A,1} (\theta,\alpha,\vec{r}_B) \\
    S_{B|A,2} (\theta,\alpha,\vec{r}_B) \\
    S_{B|A,3} (\theta,\alpha,\vec{r}_B)
\end{bmatrix} =
\sum_{kn} \text{LG}^{p=0}_{\ell_k}(\vec{r}_B) (\text{LG}^{p=0}_{\ell_n}(\vec{r}_B))^*\begin{bmatrix}
    A_{0k0n}(\theta,\alpha) + A_{1k1n}(\theta,\alpha) \\
    A_{0k1n}(\theta,\alpha) + A_{1k0n}(\theta,\alpha) \\
    i\left(A_{0k1n}(\theta,\alpha) - A_{1k0n}(\theta,\alpha)\right) \\
    A_{0k0n}(\theta,\alpha) - A_{1k1n}(\theta,\alpha)
\end{bmatrix}.
\end{equation}

\noindent An example of a set of quantum Stokes parameters for the experimentally measured state with $\ell_{1,2,3} = 0,-3,-6$ for the projection with angles $\theta = 2.36$ rad,$\alpha = $1.57 rad is shown in Fig.~\ref{fig:QSTBellSupp}e.\\

Finally, we can extract the topological structure of our tripartite state by substituting the quantum Stokes parameters into the usual skyrmion number integral,

\begin{equation}
    n(\theta, \alpha) = \frac{1}{4\pi}\int_{\mathcal{R}^2} \vec{S}_{B|A}(\theta,\alpha,\vec{r}_B) \cdot \left(\frac{\partial \vec{S}_{B|A}(\theta,\alpha,\vec{r}_B)}{\partial x} \times \frac{\partial \vec{S}_{B|A}(\theta,\alpha,\vec{r}_B)}{\partial y}\right) dx dy.
\end{equation}

We note here that the skyrmion number is now parametrized by the projection angles of photon A. Thereby yielding a new topological classification of these states, we term the "topological landscape" of the state.  Within the main text we observe two unique topological numbers per tripartite entangled state, suggesting that each state is characterized by a pair of topological numbers found by projecting photon A onto its basis states ($\theta=0,\pi$) and any superposition state ($\theta\in(0,\pi), \alpha\in[0,2\pi]$). 

\section{GHZ state extraction and topological dynamics}

\noindent In the main text we highlight an embedded GHZ-like state within our experimentally measured tripartite states and identify their characteristic Bell projection behaviour with tangible topological features and texture evolution allowing for the visualization of entanglement dynamics through topological dynamics.

\noindent To observe the embedded GHZ-like state we expand Eq.1 in the main text so that it reads
\begin{eqnarray}
    |\Psi\rangle = &&|R\rangle_A\left(|R\rangle_B |\ell_1\rangle_B + |L\rangle_B |\ell_2\rangle_B \right) + |L\rangle_A\left(|R\rangle_B |\ell_2\rangle_B + |L\rangle_B |\ell_3\rangle_B \right)
\end{eqnarray}
and then rearranging terms we find
\begin{eqnarray}
    |\Psi\rangle = &&|R\rangle_A |R\rangle_B |\ell_1\rangle_B + |L\rangle_A |L\rangle_B|\ell_3\rangle_B + \left(|R\rangle_A|L\rangle_B  + |L\rangle_A|R\rangle_B \right)|\ell_2\rangle_B, 
\end{eqnarray}
from which the GHZ-like state appears as, $|\psi^{GHZ}\rangle_{AB}  = |R\rangle_A |R\rangle_B |\ell_1\rangle_B + |L\rangle_B |L\rangle_B|\ell_3\rangle_B$ interfered with a polarization entangled reference state, $|\psi^{Ref}\rangle_{AB} = \left(|R\rangle_A|L\rangle_B  + |1\rangle_A|0\rangle_B \right)|2\rangle_B$. Given that the OAM Hilbert space of $|\psi^{GHZ}\rangle_{AB}$ is spanned by only two OAM states $\{\ket{\ell_1},\ket{\ell_3}\}$ which excludes $\ket{\ell_2}$ that multiplies the reference state, we can extract the GHZ-like state by simply projecting photon B's OAM onto the superposition $\ket{\ell_1}+\ket{\ell_3}$.\\  

\noindent As before, the topology of the state is identified post measurement of photon A's polarization. Projections onto the basis states of photon A for the GHZ state would collapse photon B into the separable states, $\ket{R}_B \ket{\ell_1}_B$ and $\ket{R}_B \ket{\ell_1}_B$, respectively, each possessing trivial topology, $n=0$. Instead projecting onto superposition states results in the non-separable states with $|n| = |\ell_3-\ell_1|$ (assuming $|\ell_3|\neq|\ell_1|$), with the topological feature. In the main text an example of a GHZ-like state extracted from the tripartite state with OAM numbers $\ell_{1,2,3} = 0,-3,-6$ results in a topological landscape with $n=0$ at the poles and $n=-6$ everywhere else. \\

\noindent We can then consider how the GHZ-like state is related to the initial tripartite state by studying the effect of the reference on the GHZ-like state. Firstly, the topological landscape is altered at the poles ($\theta = 0,\pi$) from $n=0$ to $|n| = |\ell_1 - \ell_2|$ but is left unaltered everywhere else, suggesting that the state goes through a topological transition at the poles after adding the reference but everywhere else the state is smoothly deformed. In the regime where the state undergoes a smooth deformation the central skyrmionic quasi-particle is smoothly broken up into multiple quasi-particles with skyrmion numbers that sum to give the original one. Furthermore, when changing $\alpha$ in the superposition projection the topological texture of the GHZ-like state is seen rotating, in contrast to the orbit and spin of the quasi-particle distribution seen once the reference is added. This emphasizes the nature of existent entanglement structures through the observation of evolving topological structures within the state.

\end{document}